\documentclass[twocolumn,aps,pra,showpacs,amsmath,amssymb,superscriptaddress]{revtex4}

\usepackage[dvips]{graphicx}

\begin{document}

\title{Adiabatic Hyperspherical Representation for the Three-body Problem in Two Dimensions}

\author{J. P. D'Incao}
\affiliation{JILA, University of Colorado and NIST, Boulder, Colorado 80309-0440, USA}
\affiliation{Department of Physics, Kansas State University, Manhattan, Kansas 66506, USA}
\author{B. D. Esry}
\affiliation{Department of Physics, Kansas State University, Manhattan, Kansas 66506, USA}

\begin{abstract}
We explore the three-body problem in two dimensions using the adiabatic hyperspherical
representation. We develop the main equations in terms of democratic hyperangular coordinates
and determine several symmetry properties and boundary conditions for both interacting and non-interacting solutions. 
From the analysis of the three-body effective potentials, we determine the threshold laws for low energy three-body recombination, 
collision-induced dissociation as well as inelastic atom-diatom
collisions in two dimensions. 
Our results show that the hyperspherical representation can offer a simple and
conceptually clear physical picture for three-body process in
two dimensions which is also suitable for calculations using finite range two-body interactions supporting a number of bound states. 
\end{abstract}


\maketitle

\section{Introduction}

In recent years, ultracold quantum gases have offered the most favorable
conditions for exploring universal aspects of few-body physics. Due to the experimental ability to
control interatomic interactions using Feshbach resonances \cite{chin2010RMP}, the universal few-body physics
originated by Efimov \cite{braaten2006PR,wang2013AAMOP} has become observable
and plays an important role in the stability of ultracold quantum 
gases. Other classes of universal few-body states have also emerged for a variety of 
unconventional scenarios \cite{wang2013AAMOP,wang2011PRLa,wang2011PRLb,guevara2012PRL},
expanding our knowledge of fundamental aspects of few-body systems. 

We can further expand our knowledge by 
considering physics in lower dimensions. In particular, we will consider three-body systems relevant for two-dimensional (2D) ultracold gases. 
Originally, few-body studies in reduced dimensions were motivated by the practical computational benefit. 
In the case of 2D systems, however, this simplification actually
introduces qualitatively different physical properties for the system starting at the two-body level \cite{shih2009PRA,kanjilal2006PRA} with
a strong impact on the physics of few-body systems \cite{nielsen2001PR,petrov2001PRA,liu2010PRB}.

Although the Efimov effect does not occur in 2D \cite{nielsen2001PR}, a system of three identical bosons interacting via short-range 
forces has been shown to support universal states \cite{brunch1979PRA,nielsen1999FBS,platter2004FBS,hammer2004PRL,blume2005PRB,brodsky2006PRA,
lee2006PRA,kartavtsev2006PRA,helfrich2011PRA},
and a whole new set of few-body states have now emerged \cite{bellotti2011JPB,bellotti2012PRA,bellotti2013JPB,nishida2013PRL} 
with the expectation that they will impact several properties of 2D quantum gases. Experimentally, however, it is only possible
to create a quasi-2D system, and the connection between such universal states and what can actually be observed experimentally 
is yet to be determined. Although such experimental conditions can lend themselves to novel physics
\cite{nishida2009PRA,nishida2010PRA,nishida2010PRAb,lamporesi2010PRL,haller2010PRL}, 
the study of 2D few-body systems provides important knowledge for understanding quasi-2D systems by describing their limiting behavior 
and should, in part, control the low-energy dependence of scattering observables, i.e., the threshold laws \cite{esry2001PRA,dincao2005PRL}.

In the present paper, we describe the details necessary to implement the adiabatic hyperspherical
representation for the three-body problem in 2D. This approach is general and is thus capable 
of going beyond the usual zero-range model potential used in most of the recent studies of few-body systems
in 2D \cite{kartavtsev2006PRA,helfrich2011PRA,bellotti2011JPB,
bellotti2012PRA,bellotti2013JPB,nishida2013PRL,nishida2009PRA,nishida2010PRA,nishida2010PRAb}. 
The adiabatic hyperspherical representation offers a simple and conceptually clear
description of scattering processes in terms of effective three-body potentials. Here, we explore various 
symmetry properties and derive boundary conditions that are suitable for numerical calculations 
with general interactions. Based on our analysis of the long-range behavior of the three-body potentials,
we also derive the threshold laws for the three-body scattering observables for ultracold experiments, namely,
three-body recombination, collision-induced dissociation and inelastic atom-diatom collisions. 

\section{Adiabatic Hyperspherical representation} \label{Theory}

Although different choices of hyperspherical coordinates exist \cite{lin1995PR}, the use of hyperspherical 
``democratic'', or Smith-Whitten, coordinates have proven to be extremely useful for three-atom systems 
in 3D \cite{smith1962JMP,whitten1968JMP,johnson1980JCP,lepetit1990CPL,kendrick1999JCP,suno2002PRA,
suno2010PRA,suno2008PRA,suno2009PRA}. With this coordinate system, one can conveniently describe all
fragmentation channels as well as define the boundary conditions in a manner that is
extremely beneficial for numerical implementations. 

The motion of three particles in a plane can be described by nearly the same democratic 
coordinates used in 3D and has been considered at some length by Johnson in
Ref.~\cite{johnson1983JCP}. This is possible since the first step in defining 3D democratic coordinates
is going to the body-frame, which is always a plane for three bodies.
The main difference lies in the description of the Euler angles: in 2D, we need only a single angle.  
The reduction from three Euler angles in 3D to one in 2D has consequences for the parity and permutation symmetry 
operations (as we show in App. \ref{Permutations}) that, in turn, change the hyperangular boundary conditions
from their 3D form. 

The 2D democratic coordinates are defined by first transforming
the Jacobi vectors $(\vec\rho_1,\vec\rho_2)$ in the laboratory frame (superscript $L$) to the body frame via rotation
by the Euler angle $\gamma$:
\begin{align}
\left(
 \begin{array}{cc}
  \rho_{2x}^L & \rho_{1x}^L \\
  \rho_{2y}^L & \rho_{1y}^L 
 \end{array}
\right)
& =
\left(
 \begin{array}{cc}
  \cos\gamma & -\sin\gamma \\
  \sin\gamma & \cos\gamma
 \end{array}
\right)
\left(
 \begin{array}{cc}
  \rho_{2x} & \rho_{1x} \\
  \rho_{2y} & \rho_{1y} 
 \end{array}
\right).\label{LF}
\end{align}
In 3D, Eq.~(\ref{LF}) would have $\rho_{1z}^L$ and $\rho_{2z}^L$ along with two additional Euler angles in the first matrix
on the right-hand side to make it a full 3D rotation matrix, but the body-frame coordinate matrix would not change---i.e., 
$\rho_{1z}=\rho_{2z}=0$ \cite{suno2002PRA}. In the above equation, $\vec{\rho}_1^L$ and $\vec{\rho}_2^L$ are 
``mass-scaled'' Jacobi vectors defined in terms of the individual particle's lab-frame positions $\vec r_{i}$ 
and masses $m_i$ as
\begin{eqnarray}
&&\vec{\rho}_{1}^L=(\vec{r}_{2}-\vec{r}_{1})/d_{12}, \nonumber \\
&&\vec{\rho}_{2}^L=d_{12}\left(\vec{r}_{3}
-\frac{m_{1}\vec{r}_{1}+m_{2}\vec{r}_{2}}{m_{1}+m_{2}}\right), \label{JacobiMass}
\end{eqnarray}
with
\begin{equation}
\mu d_{ij}^2 = \frac{m_k(m_i+m_j)}{m_1+m_2+m_3}
\mbox{~~and~~}\mu^2 = \frac{m_1m_2m_3}{m_1+m_2+m_3},
\end{equation}
where the indices ($i$,$j$,$k$) are a cyclic permutation of (1,2,3) and $\mu$ is the three-body reduced mass.
Note that the choice of $\mu$ is arbitrary, but the present choice ensures that the volume element remains independent of mass \cite{Delves}.

The hyperangles are defined by the condition that the moment of inertia tensor is diagonal \cite{johnson1983JCP}, 
which is accomplished by the transformation
\begin{align}
\left(
 \begin{array}{cc}
  \rho_{2x} & \rho_{1x} \\
  \rho_{2y} & \rho_{1y} 
 \end{array}
\right)& = R
\left(
 \begin{array}{cc}
  \cos\theta' &     0       \\
      0      & \sin\theta'
 \end{array}
\right)
\left(
 \begin{array}{cc}
  \cos\varphi' & \sin\varphi' \\
  -\sin\varphi' & \cos\varphi'
 \end{array}
\right).
\label{Coords}
\end{align}
Here, $\theta'=\theta/2-\pi/4$ and $\varphi'=\varphi/2+\pi/6$, with $\theta$ and $\varphi$ being the democratic hyperangles 
describing the internal motion of the particles, and 
\begin{equation}
R=(\rho_1^2+\rho_2^2)^{1/2}\label{R}
\end{equation}
 is the hyperradius
giving the overall size of the system. The hyperspherical coordinates 
are defined within the ranges
\begin{eqnarray}
\begin{array}{ccccc}
~~~~& 0\leq R < \infty,  &~~~~~~~~& & 0\leq\theta\leq\pi, \\
 & 0\leq\varphi\leq 4\pi, &~~~~~~~~&  & 0\leq\gamma\leq 2\pi.\label{Domain}
\end{array}
\end{eqnarray}
Besides the single Euler angle, the main difference between the 2D and 3D definitions
of these coordinates is the range of $\theta$.  In 3D, $\theta$ only takes on values
between 0 and $\frac{\pi}{2}$.  The change in going to 2D comes from the fact that there is no
Euler angle that can change the orientation of the body-frame $z$-axis.  In 3D, the orientation
of the $z$-axis is determined by $\vec\rho_1^L\times\vec\rho_2^L$ and is
thus a dynamical quantity. In 2D, the $z$-axis is fixed. More precisely, the plane of the particles is
fixed and thus contains all configurations of the particles---i.e., both signs of $\vec\rho_1^L\times\vec\rho_2^L$. Consequently,
the range of $\theta$ is doubled to correctly reproduce all configurations of the particles.
We will further see in App. \ref{Permutations} that $\theta$ is affected by permutations while it is not
in 3D \cite{suno2010PRA,suno2008PRA,suno2009PRA}.

With the definitions above, we can now introduce the three-body Schr\"odinger equation in 2D
for the rescaled total wave function $\Psi\rightarrow \Psi/R^{3/2}$ (atomic units will be used unless otherwise
noted)
\begin{eqnarray}
\left[-\frac{1}{2\mu}\frac{\partial^2}{\partial R^2}
+H_{\rm ad}(R,\Omega)\right]\Psi(R,\Omega)=E\Psi(R,\Omega),\label{schreq} 
\end{eqnarray} 
\noindent
where $E$ is the total energy and  $\Omega\equiv\{\theta,\varphi,\gamma\}$ denotes the set of all hyperangles. 
In Eq.~(\ref{schreq}), the adiabatic Hamiltonian is given by
\begin{eqnarray}
H_{\rm ad}(R,\Omega)=\frac{\Lambda^2(\Omega)+3/4}{2\mu R^2}
+V(R,\theta,\varphi),\label{Had}
\end{eqnarray}
containing the grand angular momentum, i.e., the hyperangular part of the kinetic energy, 
\begin{multline}
\Lambda^2(\Omega) =
-4 \left(\frac{1}{\sin\theta}\frac{\partial}{\partial\theta}\sin\theta\frac{\partial}{\partial\theta}
+\frac{1}{\sin^2\theta}\frac{\partial^2}{\partial\varphi^2} \right)\nonumber\\
-\frac{1}{\sin^2\theta}
\left( \frac{\partial^2}{\partial\gamma^2}
-4\cos\theta\frac{\partial^2}{\partial\gamma\partial\varphi}\right),
\label{2DLambda}
\end{multline}
as well as all the interparticle interactions via $V(R,\theta,\varphi)$. 
The grand angular momentum operator is essentially the same as in 3D save for a few factors of 
two \cite{suno2002PRA} and, of course, the Euler angles.

Although it is not necessary, we typically 
assume the interactions to be a pairwise sum of the form
\begin{equation}
V(R,\theta,\varphi)=v(r_{12})+v(r_{23})+v(r_{31}), \label{Int}
\end{equation} 
where the interparticle distances $r_{ij}$ are given in terms of the hyperspherical
coordinates by
\begin{eqnarray}
r_{12}&=&
2^{-1/2}{d_{12}}{R}\left[1+\sin{\theta}\cos(\varphi+\varphi_{12})\right]^{1/2}, 
\nonumber \\
r_{23}&=&
2^{-1/2}{d_{23}}{R}\left[1+\sin{\theta}\cos(\varphi+\varphi_{23})\right]^{1/2}, 
\nonumber \\
r_{31}&=&
2^{-1/2}{d_{31}}{R}\left[1+\sin{\theta}\cos(\varphi+\varphi_{31})\right]^{1/2}. \label{rij}
\end{eqnarray}
The mass-dependent angles are 
$\varphi_{12}=2\tan^{-1}(m_{3}/\mu)$, $\varphi_{23}=0$, and $\varphi_{31}=-2\tan^{-1}(m_{2}/\mu)$.
Non-additive forces be can easily introduced in Eq.~(\ref{Int}) with effectively no cost to the calculations \cite{dincao2009JPB}.

It should be noted that the wave function must satisfy the condition
\begin{equation}
\Psi(R,\theta,\varphi+2\pi,\gamma+\pi) = \Psi(R,\theta,\varphi,\gamma)
\label{OneToOne}
\end{equation}
to ensure that there is a one-to-one correspondence between these coordinates
and lab-frame coordinates [see Eqs. (\ref{LF}) and (\ref{Coords})]. 
 
In the adiabatic hyperspherical representation, the total wave 
function is expanded in terms of the channel functions
$\Phi_{\nu}(R;\Omega)$, 
\begin{equation}
\Psi(R,\Omega)=\sum_{\nu}F_{\nu}(R)\Phi_{\nu}(R;\Omega),
\label{chfun}
\end{equation}
\noindent 
where $F_{\nu}(R)$ are the hyperradial wave functions and $\nu$ represents
all quantum numbers necessary to specify each  channel. The channel functions 
$\Phi_{\nu}(R;\Omega)$ form a complete set of orthonormal functions at each value of $R$ and 
are eigenfunctions of $H_{\rm ad}$, 
\begin{equation}
H_{\rm ad}(R,\Omega)\Phi_{\nu}(R;\Omega)
=U_{\nu}(R)\Phi_{\nu}(R;\Omega).\label{poteq}
\end{equation}
\noindent
The eigenvalues $U_{\nu}(R)$ are the three-body potentials from which, as we will see next, one can define {\em effective} three-body 
potentials for the hyperradial motion. 

Substituting Eq.~(\ref{chfun}) into the Schr\"odinger equation
(\ref{schreq}) and projecting out $\Phi_{\nu'}$ (the volume element in 2D is 
$dR\sin\theta d\theta d\varphi d\gamma/4$ \cite{johnson1983JCP}), 
we obtain the hyperradial Schr\"odinger equation,  
\begin{widetext}
\begin{eqnarray}
\left[-\frac{1}{2\mu}\frac{d^2}{dR^2}+W_{\nu}(R)\right]F_{\nu}(R) 
-\frac{1}{2\mu}\sum_{\nu'\ne\nu}
\left[P_{\nu\nu'}(R)\frac{d}{dR}
+\frac{d}{dR}P_{\nu\nu'}(R)
+Q_{\nu\nu'}(R)\right]F_{\nu'}(R)
=EF_{\nu}(R),\label{radeq}
\end{eqnarray}
\end{widetext}
\noindent
that describes the motion of the three-body system under
the influence of the effective potentials 
\begin{eqnarray}
W_{\nu}(R)=U_{\nu}(R)-\frac{Q_{\nu\nu}(R)}{2\mu}. 
\end{eqnarray}
As we will see in Sec. \ref{Potentials}, including $Q_{\nu\nu}(R)$ in 
the definition of the effective potential $W_{\nu}(R)$ is crucial for obtaining potentials
with the correct behavior at large distances.
In the adiabatic hyperspherical representation, the nonadiabatic coupling terms 
$P_{\nu\nu'}(R)$ and $Q_{\nu\nu'}(R)$ ($\nu\ne\nu'$) drive inelastic collisions and are defined as    
\begin{eqnarray} 
P_{\nu\nu'}(R) &=&
\Big\langle\hspace{-0.15cm}\Big\langle\Phi_{\nu}\Big|
\frac{d}{dR}\Big|\Phi_{\nu'}\Big\rangle\hspace{-0.15cm}\Big\rangle
\label{puv}
\end{eqnarray}
\noindent
and
\begin{eqnarray} 
Q_{\nu\nu'}(R) &=&
\Big\langle\hspace{-0.15cm}\Big\langle\frac{d}{dR}\Phi_{\nu}
\Big|\frac{d}{dR}\Phi_{\nu'}\Big\rangle\hspace{-0.15cm}\Big\rangle.
\label{quv}
\end{eqnarray} 
\noindent
The double brackets denote integration over the angular
coordinates $\Omega$ only. As it stands, Eq.~(\ref{radeq}) is
exact. In practice, of course, the sum over channels must be
truncated and the number of channels retained increased until one achieves the desired accuracy.

\section{Symmetrized Hyperspherical Harmonics} \label{HypersphericalHarmonics}

To quickly assess the symmetry properties and degeneracy that are important for the large-$R$ behavior where interactions 
are negligible, we analyze the 2D non-interacting [$V(R,\theta,\varphi)=0$ in Eq.~(\ref{Had})] solutions, i.e., the
hyperspherical harmonics.
The symmetrized hyperspherical harmonics will satisfy the boundary conditions
we derive in Sec. \ref{BoundaryConditions} and thus serves as a further confirmation
of those findings. 

One interesting property of the three-body problem in 2D, unlike the 3D case, is that 
the hyperspherical harmonics can be written in closed form
\begin{eqnarray}
\Lambda^2 Y_{\omega M}^{\lambda}(\Omega)
=\lambda(\lambda+2)Y_{\omega M}^{\lambda}(\Omega),\label{EigenValueProb}
\end{eqnarray}
with $Y$ being defined in terms of the Wigner $d$-function as
\begin{equation}
Y_{\omega M}^{\lambda}(\Omega) = \frac{1}{\pi}\sqrt{\frac{\lambda+1}{4}}e^{i\frac{\omega}{2}\varphi}
d_{\frac{\omega}{2}\frac{M}{2}}^\frac{\lambda}{2}(\theta) e^{iM\gamma}.
\label{Harmonics}
\end{equation} 
Here, $\lambda$ is the hyperangular momentum quantum number,
$\omega$ is a quantum number labeling degenerate eigenstates,
and $M$ is the total orbital angular momentum.
Note that the condition in Eq.~(\ref{OneToOne}) and the fact that $-\lambda\le\{\omega,M\}\le\lambda$ 
(from the properties of the Wigner $d$-functions) imply that $\lambda$, $\omega$, and $M$ must all be either 
even or odd integers.  This condition is satisfied if these quantum numbers --- specifically $\lambda/2$, $\omega/2$ and $M/2$ ---
obey the usual rules for angular momenta if $\omega$ and $M$ are regarded as projections of $\lambda$.

With this closed form for the hyperspherical harmonics, we can explore the allowed quantum numbers for a given 
permutation symmetry when the three-body system has indistinguishable particles.
The idea is to determine the effects of the coordinate transformations 
due to parity $\Pi$ and to permutations of particles $i$ and $j$ ($P_{ij}$)
on these analytic functions. This analysis is outlined in App. \ref{Permutations} and summarized here:
\begin{align}
\Pi    Y_{\omega M}^\lambda (\Omega) &= (-)^M 
       Y_{\omega M}^\lambda (\Omega), \label{Pfirst} \\
P_{12} Y_{\omega M}^\lambda (\Omega) &= (-)^\frac{3M+\lambda}{2} e^{i\omega\frac{2\pi}{3}}
       Y_{-\omega M}^\lambda (\Omega), \\
P_{23} Y_{\omega M}^\lambda (\Omega) &= (-)^\frac{M+\lambda}{2} e^{i\omega\frac{\pi}{3}}
       Y_{-\omega M}^\lambda (\Omega), \\
P_{31} Y_{\omega M}^\lambda (\Omega) &= (-)^\frac{3M+\lambda}{2}
       Y_{-\omega M}^\lambda (\Omega), \\
P_{12}P_{23} Y_{\omega M}^\lambda (\Omega) &= (-)^M e^{i\omega\frac{\pi}{3}}
       Y_{\omega M}^\lambda (\Omega), \\
P_{12}P_{31} Y_{\omega M}^\lambda (\Omega) &= e^{i\omega\frac{2\pi}{3}}
       Y_{\omega M}^\lambda (\Omega). \label{Plast}
\end{align}
Note that the hyperradius is invariant under all the symmetry operations above.
Note also that in the derivation of Eqs. (\ref{Pfirst})--(\ref{Plast}) the only non-trivial
relation used was that the $d$-function in Eq.~(\ref{Harmonics}) has the following property:
$d_{mm'}^\ell (\pi-\theta) = (-)^{\ell+m'}d_{-mm'}^\ell(\theta)$.
As one can see, the hyperspherical harmonics are already parity eigenstates. 
Since $M$ is the total orbital angular momentum and is thus a good quantum number, 
none of the permutation operators change $M$. 
The symmetrized harmonics we will construct based on Eqs. (\ref{Pfirst})--(\ref{Plast}) will thus
remain parity eigenstates.

\subsection{Three identical bosons}

For three identical bosons ($BBB$), we are interested in the completely symmetric hyperspherical 
harmonics. To obtain these states, we apply the (un-normalized) symmetrization operator,
${\cal S}=(1+P_{12}+P_{23}+P_{31}+P_{12}P_{23}+P_{12}P_{31})$, to the hyperspherical harmonics defined in Eq.~(\ref{Harmonics})
and use Eqs.~(\ref{Pfirst})--(\ref{Plast}). Doing so, we find
\begin{align}
{\cal S} Y_{\omega M}^{\lambda}(\Omega)  &= \left(1+(-)^M e^{i\omega\frac{\pi}{3}}+e^{i\omega\frac{2\pi}{3}}\right)\nonumber \\
  &  ~~~~\times\left(Y_{\omega M}^{\lambda}(\Omega) +(-)^\frac{3M+\lambda}{2} 
                          Y_{-\omega M}^{\lambda}(\Omega) \right).
\label{ThreeBosonFn}
\end{align}
The prefactor here should not vanish. This condition determines, for a given value of $M$, the allowed values for $\lambda$ and $\omega$.
By inspection, we find that we must have
\begin{equation}
\omega =
\begin{cases}
6n, & ~M~{\rm even} \\
6n+3, & ~ M~ {\rm odd}.
\end{cases}
~~(n=0,1,2,...)\label{nuCond1}
\end{equation}
Note that these conditions are essentially what we found for the 3D case \cite{suno2002PRA}.
Now, we recall from Eq.~(\ref{OneToOne}) that $\lambda$, $\omega$, and $M$ must all
be even or odd.  Equation~(\ref{ThreeBosonFn}) further tells us that if $\frac{3M+\lambda}{2}$
is odd, then $\omega$ cannot be zero or the function will vanish.  But, $\omega$ can only
be zero if $M$ is even.  So, we amend Eq.~(\ref{nuCond1}) to
\begin{equation}
\omega =
\begin{cases}
6n, & (n\neq 0~{\rm if}~\frac{3M+\lambda}{2}{\rm ~odd)},~M~{\rm even} \\
6n+3, &~ M~ {\rm odd}.
\end{cases}
\end{equation}
Explicit examples for the allowed quantum numbers are given in Table~\ref{ThreeBosonTab} 
for the lowest few values of $M$. 

\begin{table}[htbp]
\begin{ruledtabular}
\caption{Sample of allowed hyperspherical harmonic quantum numbers for three identical bosons ($BBB$).
We list the lowest few values for both $\lambda$ and $M$.}
\begin{tabular}{cccccccc}
  \multicolumn{2}{c}{$M=0$} & \multicolumn{2}{c}{$|M|=1$}  & \multicolumn{2}{c}{$|M|=2$} & \multicolumn{2}{c}{$|M|=3$}\\ 
 $\lambda$ &  $|\omega|$ &  $\lambda$ &  $|\omega|$&  $\lambda$ &  $|\omega|$ &  $\lambda$ &  $|\omega|$ \\ \hline
 0 & 0 &  3 & 3 &  2 & 0 & 3 & 3 \\
 4 & 0 &  5 & 3 &  6 & 0,6 & 5 & 3 \\
 6 & 6 &  7 & 3 &  8 & 6 & 7 & 3 \\
 8 & 0,6 &  9 & 3,9 &  10 & 0,6 & 9 & 3,9 
\end{tabular}
\label{ThreeBosonTab}
\end{ruledtabular}
\end{table}

\subsection{Three identical fermions}

For three identical fermions ($FFF$), we are interested in the completely antisymmetric hyperspherical harmonics.
So, we now apply the anti-symmetrization operator, ${\cal A} =   (1-P_{12}-P_{23}-P_{31}+P_{12}P_{23}+P_{12}P_{31})$, to Eq.~(\ref{Harmonics}) and 
find
\begin{align}
{\cal A} Y_{\omega M}^{\lambda}(\Omega) &= \left(1+(-)^M e^{i\omega\frac{\pi}{3}}+e^{i\omega\frac{2\pi}{3}}\right) \nonumber\\
   &~~~~ \left(Y_{\omega M}^{\lambda}(\Omega)-(-)^\frac{3M+\lambda}{2} 
                          Y_{-\omega M}^{\lambda}(\Omega) \right).
\label{ThreeFermionFn}
\end{align}
Just as for the bosons, the prefactor here should not vanish, so we must have
\begin{equation}
\omega =
\begin{cases}
6n, & M~{\rm even} \\
6n+3, & M~ {\rm odd}. \label{nuCond2}
\end{cases}
\end{equation}
These conditions are exactly the same as for bosons. The difference due to antisymmetry stems from the fact that $\frac{3M+\lambda}{2}$
must be odd when $\omega=0$ or the function will vanish (opposite the case for bosons).  
But, $\omega$ can only be zero if $M$ is even, so $\lambda/2$ must be {\it odd}. 
So, we qualify Eq.~(\ref{nuCond2}) as
\begin{equation}
\omega =
\begin{cases}
6n, & (n\neq 0~{\rm if}~\frac{3M+\lambda}{2}{\rm ~even}),~M~{\rm even} \\
6n+3, &~ M~ {\rm odd}.
\end{cases}
\end{equation}
Explicit examples are given in Table~\ref{ThreeFermionTab}.

\begin{table}[htbp]
\begin{ruledtabular}
\caption{Same as Table \ref{ThreeBosonTab} but for three identical fermions ($FFF$).}
\begin{tabular}{cccccccc}
\multicolumn{2}{c}{$M=0$} & \multicolumn{2}{c}{$|M|=1$} & \multicolumn{2}{c}{$|M|=2$} & \multicolumn{2}{c}{$|M|=3$}\\ 
  $\lambda$ &  $|\omega|$ &  $\lambda$ &  $|\omega|$ &  $\lambda$ &  $|\omega|$ &  $\lambda$ &  $|\omega|$ \\ \hline
  2 & 0 &  3 & 3 &  4 & 0 & 3 & 3 \\
  6 & 0,6 &  5 & 3 &  6 & 6 & 5 & 3 \\
  8 & 6 &  7 & 3 &  8 & 0,6 & 7 & 3 \\
  10 & 0,6 &  9 & 3,9 &  10 & 6 & 9 & 3,9  
\end{tabular}
\label{ThreeFermionTab}
\end{ruledtabular}
\end{table}

\subsection{Two identical bosons}

When there are only two identical bosons ($BBX$) and they are labeled 1 and 3, we symmetrize the hyperspherical harmonics by applying
${\cal S}= (1+P_{31})$ to Eq.~(\ref{Harmonics}). In this case, symmetrization for bosons requires
\begin{align}
{\cal S} Y_{\omega M}^{\lambda}(\Omega) 
 &= Y_{\omega M}^{\lambda}(\Omega)+(-)^\frac{3M+\lambda}{2} 
                          Y_{-\omega M}^{\lambda}(\Omega).
\label{TwoBosonFn}
\end{align}
There are thus no restrictions on $\omega$ except that $\omega\neq 0$ if $\frac{3M+\lambda}{2}$
is odd.  Some of the allowed $\lambda$ and $\omega$ are given in Table~\ref{TwoBosonTab}.

\begin{table}[htbp]
\begin{ruledtabular}
\caption{Same as Table \ref{ThreeBosonTab} but for two identical bosons ($BBX$).}
\begin{tabular}{cccccccc}
\multicolumn{2}{c}{$M=0$} & \multicolumn{2}{c}{$|M|=1$}& \multicolumn{2}{c}{$|M|=2$}& \multicolumn{2}{c}{$|M|=3$} \\
  $\lambda$ &  $|\omega|$ & $\lambda$ & $|\omega|$& $\lambda$ & $|\omega|$& $\lambda$ & $|\omega|$ \\ \hline
  0 & 0 &         1 & 1 &                        2 & 0,2 &3 & 1,3 \\
  2 & 2 &         3 & 1,3 &                 4 & 2,4 & 5 & 1,3,5\\
  4 & 0,2,4 &    5 & 1,3,5 &          6 & 0,2,4,6 & 7 & 1,3,5,7\\
  6 & 2,4,6 &  7 & 1,3,5,7 &   8 & 2,4,6,8 & 9 & 1,3,5,7,9
\end{tabular}
\label{TwoBosonTab}
\end{ruledtabular}
\end{table}

\subsection{Two identical fermions}

To complete our analysis, we consider the case of two identical fermions ($FFX$), i.e. applying ${\cal A}=(1-P_{31})$ to Eq.~(\ref{Harmonics}). For this case, we find
\begin{align}
{\cal A} Y_{\omega M}^{\lambda}(\Omega) 
 &= Y_{\omega M}^{\lambda}(\Omega)-(-)^\frac{3M+\lambda}{2} 
                          Y_{-\omega M}^{\lambda}(\Omega).
\label{TwoFermionFn}
\end{align}
Again, there are no restrictions on $\omega$ except that $\omega\neq 0$ if $\frac{3M+\lambda}{2}$
is now even.  Some of the allowed $\lambda$ and $\omega$ are given in Table~\ref{TwoFermionTab}.

\begin{table}[htbp]
\begin{ruledtabular}
\caption{Same as Table \ref{ThreeBosonTab} but for two identical fermions ($FFX$).}
\begin{tabular}{cccccccc}
\multicolumn{2}{c}{$M=0$} & \multicolumn{2}{c}{$|M|=1$}& \multicolumn{2}{c}{$|M|=2$} & \multicolumn{2}{c}{$|M|=3$} \\ 
  $\lambda$ & $|\omega|$ & $\lambda$ & $|\omega|$& $\lambda$ & $|\omega|$ & $\lambda$ & $|\omega|$ \\ \hline
  2 & 0,2 &         1 & 1 &                        2 & 2 & 3 &1,3\\
  4 & 2,4 &       3 & 1,3 &                 4 & 0,2,4 & 5 & 1,3,5\\
  6 & 0,2,4,6 &    5 & 1,3,5 &          6 & 2,4,6 & 7 & 1,3,5,7\\
  8 & 2,4,6,8 &  7 & 1,3,5,7 &   8 & 0,2,4,6,8 & 9 & 1,3,5,7,9
\end{tabular}
\label{TwoFermionTab}
\end{ruledtabular}
\end{table}

\section{Boundary conditions} \label{BoundaryConditions}

While the symmetrized harmonics of Sec. \ref{HypersphericalHarmonics} can be used as a basis to expand $\Phi_{\nu}$
and solve Eq.~(\ref{poteq}), they are an inefficient choice in practice. The difficulty with this basis is due to the localization
of $\Phi_{\nu}$ in the hyperangular plane as $R$ increases, requiring the number of basis functions to grow.
Specifically, a simple uncertainty argument shows that their number must grow at least linearly with $R$ in order
to describe the localized two-body channels.
More flexible methods such as b-splines or finite elements have proven
much more effective \cite{suno2002PRA}. Symmetries must still be imposed, however, and can actually improve these methods' efficiency 
by reducing the required integration domain. This reduction is accomplished by imposing boundary conditions on the smallest unique region
the symmetry allows. For instance, three identical particles permits a reduction of the integration domain by $3!$ \cite{suno2002PRA}.

In this section we will derive these boundary conditions.
Although the analysis in the previous section already gives the information necessary 
for obtaining the boundary conditions, we will take an alternative, independent approach here.
The results, of course, are equivalent.
Note that since the hyperradius is invariant under symmetry
operations in Eqs.~(\ref{Pfirst})--(\ref{Plast}), symmetry-motivated boundary conditions need only be imposed in the 
hyperangles. 

For isotropic two-body interactions, such that $M$ is a good quantum number, the channel functions
are separable in the Euler angle $\gamma$, with the corresponding solution normally expressed by $\exp(i M\gamma)$. 
As a result, $\Phi_{\nu}$ is an eigenstate of $L_{z}=i\partial/\partial\gamma$.
In this basis, however, the adiabatic Schr\"odinger equation, Eq.~(\ref{poteq}), is complex and the boundary 
conditions are difficult to implement. As we will show below, a change of basis to $\sin (M\gamma)$ and $\cos (M\gamma)$, 
\begin{eqnarray}
\Phi(R;\Omega)&=&\phi_{s}(R;\theta,\varphi)\sin M\gamma\nonumber\\
&&~~~~~~~~~+\phi_{c}(R;\theta,\varphi)\cos M\gamma,\label{PhiSC}
\end{eqnarray}
transforms the adiabatic equation into a real system of equations,
\begin{align} 
&\left[-4 \left(\frac{1}{\sin\theta}\frac{\partial}{\partial\theta}\sin\theta\frac{\partial}{\partial\theta}
+\frac{1}{\sin^2\theta}\frac{\partial^2}{\partial\varphi^2} \right)+\frac{M^2}{\sin^2\theta}\right]
\left(\hspace{-0.05in}\begin{array}{c}\phi_s \\ \phi_c\end{array}\hspace{-0.05in}\right)
\nonumber\\
&+4\frac{\cos\theta}{\sin^2\theta}M
\frac{\partial}{\partial\varphi} \left(\hspace{-0.05in}\begin{array}{c}-\phi_c \\ \phi_s\end{array}\hspace{-0.05in}\right)
=2\mu R^2U(R) \left(\hspace{-0.05in}\begin{array}{c}\phi_s \\ \phi_c\end{array}\hspace{-0.05in}\right),\label{HadSyst}
\end{align} 
where each component $\phi_s$ and $\phi_c$ will have its own set of boundary conditions, making this basis
more convenient computationally. This choice, of course, also implies that $\Phi_{\nu}$ is an eigenstate of $L_{z}^2$ 
rather than of $L_z$.

\subsection{Reflection Symmetry}

We start the present analysis by first noting that, besides the symmetry operations in Eqs.~(\ref{Pfirst})--(\ref{Plast}) (see also App. 
\ref{Permutations}), $H_{\rm ad}$ is invariant under the operations
\begin{eqnarray}
&& R_x:~(\theta,\varphi,\gamma)\rightarrow (\pi-\theta,\varphi,2\pi-\gamma),\label{Rx}\\
&& R_y:~(\theta,\varphi,\gamma)\rightarrow (\pi-\theta,\varphi,\pi-\gamma),\label{Ry}
\end{eqnarray}
where $R_x$ and $R_y$ are reflections along the body-frame $x$ and $y$ axes, respectively, as can be verified by
applying the above transformations in Eq.~(\ref{Coords}). Since $ H_{\rm ad}$ commutes with $R_x$ and $R_y$, they can share common eigenstates.
Therefore, the solutions of $ H_{\rm ad}$, i.e., the channel functions $\Phi_\nu$ in Eq.~(\ref{poteq}),
can be chosen to obey the boundary conditions resulting from these reflections.
We note, however, that since $ R_{x} R_y=\Pi$ we only need to specify the boundary conditions with respect to 
one of the reflections, which we arbitrarily choose to be $ R_{x}$. 
We also note that the $\sin (M\gamma)$ and $\cos (M\gamma)$ functions in Eq.~(\ref{PhiSC}) are eigenstates of $ R_{x}$ ($ R_{y}$) 
with eigenvalues $-1$($+1$) and $+1$($-1$), respectively, while $\exp(iM\gamma)$ is not.
Expressing $\Phi_{\nu}$ as in Eq.~(\ref{PhiSC}), we have thus chosen to construct eigenstates of $\{H_{\rm ad}, \Pi,L_z^2,R_x\}$
instead of $\{H_{\rm ad}, \Pi,L_z\}$, purely for our computational convenience. Three-body states are thus labeled by $|M|^{\pi}_r$ where
$r$ indicates the $R_x$ symmetry as defined below.

Now, in order to establish the boundary conditions due to $R_x$, we recognize 
that the channel functions, irrespective of the system's permutation symmetry, can only
be symmetric or antisymmetric with respect to reflections. This allows us to write, 
\begin{eqnarray}
 R_{x} \Phi(R;\Omega) = (-)^r \Phi(R;\Omega), \label{RxBC}
\end{eqnarray}
where $r=0$ for symmetric solutions and $r=1$ for antisymmetric solutions. Moreover, since
$ R_{x}$ [Eq.~(\ref{Rx})] keeps $\varphi$ unchanged, the boundary conditions due to reflections
will only affect $\theta$. Substituting Eq. (\ref{PhiSC}) into 
Eq. (\ref{RxBC}) and projecting out $\sin (M\gamma)$ and $\cos (M\gamma)$, we obtain
\begin{eqnarray}
\phi_s(R;\pi-\theta,\varphi)&=&(-)^{r+1} \phi_s(R;\theta,\varphi),\label{RxS}  \\
\phi_c(R;\pi-\theta,\varphi)&=&(-)^{r} \phi_c(R;\theta,\varphi),\label{RxC}
\end{eqnarray} 
and
\begin{eqnarray}
\frac{\partial}{\partial \theta}\phi_s(R;\pi-\theta,\varphi)&=&(-)^{r} \frac{\partial}{\partial \theta}\phi_s(R;\theta,\varphi),\label{RxdS} \\
\frac{\partial}{\partial \theta}\phi_c(R;\pi-\theta,\varphi)&=&(-)^{r+1} \frac{\partial}{\partial \theta}\phi_c(R;\theta,\varphi). \label{RxdC}
\end{eqnarray} 

Equations (\ref{RxS})--(\ref{RxdC}) imply that $\phi_s$ and $\phi_c$ are either even or odd upon reflection through $\theta=\pi/2$. We
can thus extract the following boundary conditions at $\theta=\pi/2$. For symmetric solutions ($r=0$),
\begin{eqnarray}
\phi_s(R,\frac{\pi}{2},\varphi)=0~~\mbox{and}~~\frac{\partial}{\partial \theta}\phi_c(R;\theta,\varphi)\Big|_{\theta=\frac{\pi}{2}}=0.
\end{eqnarray}
For antisymmetric solutions ($r=1$), the boundary conditions are:
\begin{eqnarray}
\frac{\partial}{\partial \theta}\phi_s(R;\theta,\varphi)\Big|_{\theta=\frac{\pi}{2}}=0,~~\mbox{and}~~\phi_c(R,\frac{\pi}{2},\varphi)=0.
\end{eqnarray}
Since the volume element vanishes at $\theta=0$, no boundary condition is required.
The integration domain has thus been reduced to $0\le\theta\le\pi/2$, thereby reducing the numerical effort by a factor of 2. 

\subsection{Permutation Symmetry}

The boundary conditions in $\varphi$ originate from the permutation symmetry, i.e., they will
depend on the bosonic or fermionic character of the three particles. As we will see, however, the reflection
symmetry [Eqs. (\ref{RxS}) and (\ref{RxC})], as well as the condition in Eq.~(\ref{OneToOne}), will also be used in deriving
the boundary conditions in $\varphi$.

By assuming the atoms are in a spin-stretched state, i.e., the spin part of the wave function is symmetric under permutations, 
the bosonic or fermionic symmetry must be satisfied by the spatial part of the wave function. In fact,
the boundary conditions for the channel functions are derived from the requirement that they must have a well-defined 
permutation symmetry, namely,
\begin{eqnarray}
P_{ij}\Phi(R;\Omega) = (-)^s\Phi(R;\Omega). \label{PermSym}
\end{eqnarray}
In the above equation, $s=0$ for a pair of identical bosons while $s=1$ for identical fermions. Two dissimilar 
particles require no such condition. Otherwise, the derivation of the boundary conditions in $\varphi$ follows closely,
in spirit, to the $\theta$-boundary-condition derivation in the previous section.

\subsubsection{Three identical particles}\label{3IP}

Although there exists a total of five possible permutations of three particles, i.e.,
$P_{12}$, $P_{23}$, $P_{31}$, $P_{12}P_{23}$, and $P_{12}P_{31}$, we only need to symmetrize with respect to
two of them (see App. \ref{Permutations}).
This simplification follows from the fact that the $S_{3}$ permutation group has only two generators so that all permutations
can be written as products of any two permutations \cite{Hamermesh}. We will, therefore, choose $P_{31}$ and $P_{23}$
as generators and derive the boundary conditions based on these operations.

Since we expect a $3!$ reduction in the integration domain, we will seek boundary conditions at $\varphi=0$ and
$\varphi=\pi/3$ [Eq.~(\ref{OneToOne}) already reduced the range of $\varphi$ from Eq.~(\ref{Domain}) to $0\le\varphi\le2\pi$].
So, substituting $\Phi$ from Eq. (\ref{PhiSC}) into Eq. (\ref{PermSym}), using
$P_{31}(\theta,\varphi,\gamma)=(\pi-\theta,2\pi-\varphi,\gamma)$ from Eq.~(\ref{P31}),
and projecting out $\sin (M\gamma)$ and 
$\cos (M\gamma)$, we obtain at $\varphi=0$
\begin{eqnarray}
\phi_s(R;\theta,0)&=&(-)^{r+s+M+1} \phi_s(R;\theta,0),\label{P31S}  \\
\phi_c(R;\theta,0)&=&(-)^{r+s+M}\phi_c(R;\theta,0),\label{P31C}
\end{eqnarray} 
and
\begin{align}
&\frac{\partial}{\partial \varphi}\phi_s(R;\theta,\varphi) \Big|_{\varphi=0}=(-)^{r+s+M} \frac{\partial}{\partial \varphi}\phi_s(R;\theta,\varphi) \Big|_{\varphi=0},\label{P31dS} \\
&\frac{\partial}{\partial \varphi}\phi_c(R;\theta,\varphi) \Big|_{\varphi=0}=(-)^{r+s+M+1} \frac{\partial}{\partial \varphi}\phi_c(R;\theta,\varphi) \Big|_{\varphi=0}. \label{P31dC}
\end{align}
We note that to obtain Eqs.~(\ref{P31S})--(\ref{P31dC}) we used Eqs. (\ref{RxS}) and (\ref{RxC}) which introduced the dependence on $r$,
as well as Eq. (\ref{OneToOne}), which introduced the dependence on $M$. 

For $\varphi=\pi/3$,
we similarly use Eqs.~(\ref{P23a}) and (\ref{P23b}), $P_{23}(\theta,\varphi,\gamma)=(\pi-\theta,2\pi/3-\varphi,\gamma)$, to obtain 
the following conditions
\begin{eqnarray}
\phi_s(R;\theta,\frac{\pi}{3})&=&(-)^{r+s+1} \phi_s(R;\theta,\frac{\pi}{3}),\label{P23S}  \\
\phi_c(R;\theta,\frac{\pi}{3})&=&(-)^{r+s}\phi_c(R;\theta,\frac{\pi}{3}),\label{P23C}
\end{eqnarray} 
and 
\begin{align}
&\frac{\partial}{\partial \varphi}\phi_s(R;\theta,\varphi) \Big|_{\varphi=\frac{\pi}{3}}=(-)^{r+s} \frac{\partial}{\partial \varphi}\phi_s(R;\theta,\varphi) \Big|_{\varphi=\frac{\pi}{3}},\label{P23dS} \\
&\frac{\partial}{\partial \varphi}\phi_c(R;\theta,\varphi) \Big|_{\varphi=\frac{\pi}{3}}=(-)^{r+s+1} \frac{\partial}{\partial \varphi}\phi_c(R;\theta,\varphi) \Big|_{\varphi=\frac{\pi}{3}}. \label{P23dC}
\end{align}
Therefore, we need only integrate Eq.~(\ref{poteq})
from $\varphi=0$ to $\varphi=\pi/3$, making the integration domain a factor of six smaller.
Note that, similar to the boundary conditions in $\theta$, Eqs.~(\ref{P23S})--(\ref{P23dC}) finally only require 
$\phi_s$ and $\phi_c$---or their derivative---to vanish and are thus relatively simple to implement in practice.

\subsubsection{Two identical particles}

For systems with only two identical particles, the determination of the boundary conditions is simpler
since the channel function must be symmetrized for only a single permutation. Here, we will assume
that particles 1 and 3 are identical such that the relevant permutation is $P_{31}$, leading us to seek boundary conditions at $\varphi=0$ and 
$\varphi=\pi$. 
Therefore, proceeding as in Sec. \ref{3IP}, substituting $\Phi$ into (\ref{PermSym}), where $P_{31}(\theta,\varphi,\gamma)=(\pi-\theta,2\pi-\varphi,\gamma)$ from Eq.~(\ref{P31}),
and projecting out $\sin (M\gamma)$ and $\cos (M\gamma)$, we obtain at $\varphi=0$
\begin{eqnarray}
\phi_s(R;\theta,0)&=&(-)^{r+s+M+1} \phi_s(R;\theta,0),\label{P31Sx}  \\
\phi_c(R;\theta,0)&=&(-)^{r+s+M}\phi_c(R;\theta,0),\label{P31Cx}
\end{eqnarray} 
and 
\begin{align}
&\frac{\partial}{\partial \varphi}\phi_s(R;\theta,\varphi) \Big|_{\varphi=0}=(-)^{r+s+M} \frac{\partial}{\partial \varphi}\phi_s(R;\theta,\varphi) \Big|_{\varphi=0},\label{P31dSx} \\
&\frac{\partial}{\partial \varphi}\phi_c(R;\theta,\varphi) \Big|_{\varphi=0}=(-)^{r+s+M+1} \frac{\partial}{\partial \varphi}\phi_c(R;\theta,\varphi) \Big|_{\varphi=0}. \label{P31dCx}
\end{align}
These boundary conditions are identical to Eqs.~(\ref{P31S})--(\ref{P31dC}).

To obtain the other set of boundary conditions, we repeat the process above without
imposing the condition in Eq. (\ref{OneToOne}). This yields the following conditions at $\varphi=\pi$
\begin{eqnarray}
\phi_s(R;\theta,{\pi})&=&(-)^{r+s+1} \phi_s(R;\theta,{\pi}),\label{P23Sx}  \\
\phi_c(R;\theta,{\pi})&=&(-)^{r+s}\phi_c(R;\theta,{\pi}),\label{P23Cx}
\end{eqnarray} 
and 
\begin{align}
&\frac{\partial}{\partial \varphi}\phi_s(R;\theta,\varphi) \Big|_{\varphi=\pi}=(-)^{r+s} \frac{\partial}{\partial \varphi}\phi_s(R;\theta,\varphi) \Big|_{\varphi=\pi},\label{P23dSx} \\
&\frac{\partial}{\partial \varphi}\phi_c(R;\theta,\varphi) \Big|_{\varphi=\pi}=(-)^{r+s+1} \frac{\partial}{\partial \varphi}\phi_c(R;\theta,\varphi) \Big|_{\varphi=\pi}. \label{P23dCx}
\end{align}
The boundary conditions above have the same form as Eqs.~(\ref{P23S})--(\ref{P23dC}) but are evaluated at $\varphi=\pi$ instead of $\pi/3$.
Therefore, for systems with two identical particles we need only to integrate Eq.~(\ref{poteq})
from $\varphi=0$ to $\varphi=\pi$, reducing the integration domain by a factor of 2. 

In Table \ref{TabBC}, we summarize all the boundary conditions derived in this section for systems with three identical particles ($BBB$ and $FFF$) as well
as for systems with two identical particles ($BBX$ and $FFX$). 
\begin{table}[htbp]
\begin{ruledtabular}
\caption{Summary of the boundary conditions for bosonic ($BBB$ and $BBX$) and fermionic ($FFF$ and $FFX$) systems. The table
indicates the relevant quantum number for permutation ($s$) and reflection ($r$) symmetry. The boundary conditions at $\theta=0$, $\varphi=0$,
and $\varphi=\pi/3$ (for $BBB$ and $FFF$) or $\pi$ (for $BBX$ and $FFX$) specify whether the components $\phi_s$ and $\phi_c$ of $\Phi$, or their
derivative, vanishes ($\partial_\theta=\partial/\partial\theta$ and $\partial_\varphi=\partial/\partial\varphi$).}
\begin{tabular}{ccccc}
\multicolumn{5}{l}{$BBB$ or $BBX$ ($s=0$)} \\
$r$ & $M$ & $\theta=\pi/2$ & $\varphi=0$ & $\varphi=\pi/3$ or $\pi$  \\ \hline
 0   & even & $\{\phi_s$$,$$\partial_\theta\phi_c\}$$=$$0$ & $\{\phi_s$$,$$\partial_\varphi\phi_c\}$$=$$0$ & $\{\phi_s$$,$$\partial_\varphi\phi_c\}$$=$$0$ \\
      & odd  & $\{\phi_s$$,$$\partial_\theta\phi_c\}$$=$$0$ & $\{\partial_\varphi\phi_s$$,$$\phi_c\}$$=$$0$ & $\{\phi_s$$,$$\partial_\varphi\phi_c\}$$=$$0$ \\[0.05in]
 1   & even & $\{\partial_\theta\phi_s$$,$$\phi_c\}$$=$$0$ & $\{\partial_\varphi\phi_s$$,$$\phi_c\}$$=$$0$ & $\{\partial_\varphi\phi_s$$,$$\phi_c\}$$=$$0$ \\
      & odd  & $\{\partial_\theta\phi_s$$,$$\phi_c\}$$=$$0$ & $\{\phi_s$$,$$\partial_\varphi\phi_c\}$$=$$0$ & $\{\partial_\varphi\phi_s$$,$$\phi_c\}$$=$$0$ \\\hline
\multicolumn{5}{l}{$FFF$ or $FFX$ ($s=1$)} \\
$r$ & $M$ & $\theta=\pi/2$ & $\varphi=0$ & $\varphi=\pi/3$ or $\pi$  \\ \hline
 0   & even & $\{\phi_s$$,$$\partial_\theta\phi_c\}$$=$$0$ & $\{\partial_\varphi\phi_s$$,$$\phi_c\}$$=$$0$ & $\{\partial_\varphi\phi_s$$,$$\phi_c\}$$=$$0$ \\
      & odd  & $\{\phi_s$$,$$\partial_\theta\phi_c\}$$=$$0$ & $\{\phi_s$$,$$\partial_\varphi\phi_c\}$$=$$0$ & $\{\partial_\varphi\phi_s$$,$$\phi_c\}$$=$$0$ \\[0.05in]
 1   & even & $\{\partial_\theta\phi_s$$,$$\phi_c\}$$=$$0$ & $\{\phi_s$$,$$\partial_\varphi\phi_c\}$$=$$0$ & $\{\phi_s$$,$$\partial_\varphi\phi_c\}$$=$$0$ \\
      & odd  & $\{\partial_\theta\phi_s$$,$$\phi_c\}$$=$$0$ & $\{\partial_\varphi\phi_s$$,$$\phi_c\}$$=$$0$ & $\{\phi_s$$,$$\partial_\varphi\phi_c\}$$=$$0$ \\
\end{tabular}
\label{TabBC}
\end{ruledtabular}
\end{table}

\subsubsection{Three distinguishable particles}

For three distinguishable particles, of course, no permutational symmetry is required, resulting in no boundary conditions
in $\varphi$. The only boundary condition in $\varphi$ is the one provided by Eq.~(\ref{OneToOne}) for the
one-to-one correspondence of the wave function in the body- and lab-frames.
This results in periodic boundary conditions for the channel functions, and their derivative, given by
\begin{eqnarray}
&\phi_s(R;\theta,0)=(-)^{M} \phi_s(R;\theta,2\pi),  \\
&\phi_c(R;\theta,0)=(-)^{M}\phi_c(R;\theta,2\pi),\\
&\frac{\partial}{\partial \varphi}\phi_s(R;\theta,\varphi) \Big|_{\varphi=0}=(-)^{M+1} \frac{\partial}{\partial \varphi}\phi_s(R;\theta,\varphi) \Big|_{\varphi=2\pi}, \\
&\frac{\partial}{\partial \varphi}\phi_c(R;\theta,\varphi) \Big|_{\varphi=0}=(-)^{M+1} \frac{\partial}{\partial \varphi}\phi_c(R;\theta,\varphi) \Big|_{\varphi=2\pi}.
\end{eqnarray}
For this case, therefore, the only reduction in the integration domain in Eq.~(\ref{Domain}) is provided
by the reflection symmetry which reduces the range in $\theta$ by a factor of 2.

\section{Structure of the three-body potentials in 2D} \label{Potentials}

One of the advantages of the adiabatic hyperspherical representation is that besides producing
numerically accurate results it also offers a simple and conceptually clear description of the system
in terms of the three-body potentials, $U_{\nu}(R)$. The three-body potentials are
obtained by solving the adiabatic equation (\ref{poteq}) for fixed values of $R$. Once this step is completed,
one can solve the hyperradial  Schr\"odinger equation (\ref{radeq}) to obtain the three-body bound and scattering states
from which any three-body observable can be computed. 

\begin{figure*}[htbp]
\includegraphics[width=6.9in]{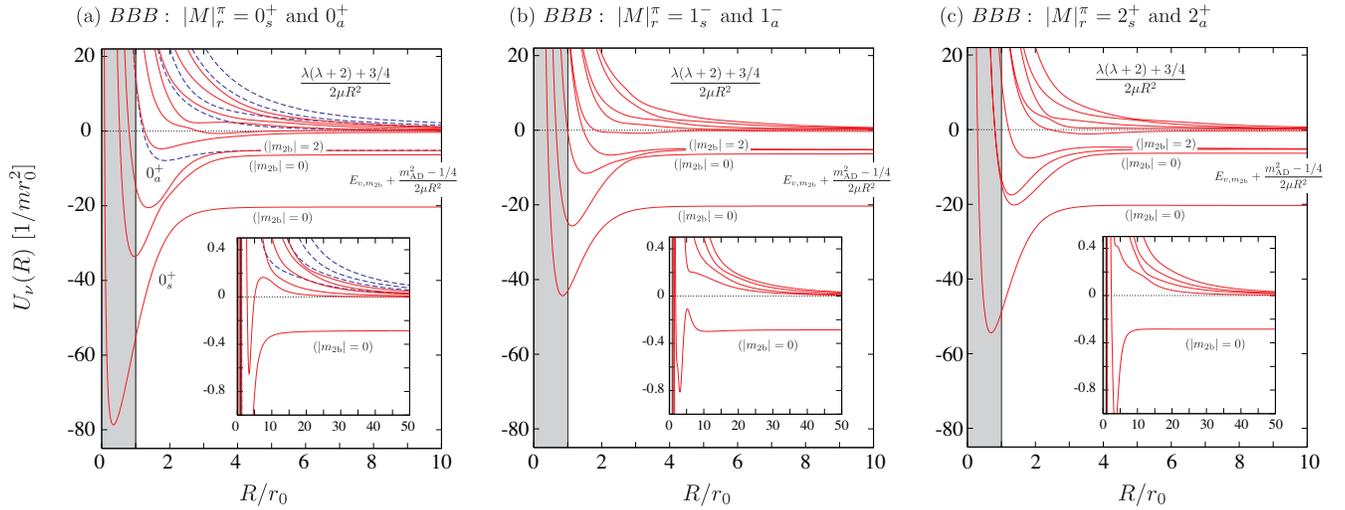}
\caption{(color online) Three-body potentials for three identical bosons in 2D with (a) $|M|^{\pi}_{r}=0^+_s$ and $0^+_a$,
(b) $|M|^{\pi}_{r}=1^-_s$ and $1^-_a$, and (c) $|M|^{\pi}_{r}=2^+_s$ and $2^+_a$ (shaded region indicates the region in $R$ in which particles can be found
at distances smaller than the range of the interaction, $r_0$). Note that for $|M|\ne0$ the
potentials for the symmetric and antisymmetric three-body systems (with respect to reflection) are exactly degenerate,
while for $|M|=0$ they are not. Note also that $m_{\rm 2b}=0$ states are forbidden for the $0^+_{a}$ symmetry in (a).}\label{BosonsFig}
\end{figure*}

\subsection{Numerical Details}

For 2D three-body systems, we must thus solve the 
two coupled partial differential equations in $\theta$ and $\varphi$ (one for $M=0$) in Eq.~(\ref{HadSyst}).
The resulting differential equations are solved by expanding $\phi_{s}$ and $\phi_c$ in Eq.~(\ref{PhiSC}) onto a direct product of basis splines 
\cite{Splines}, in the same spirit as the 3D three-body problem \cite{suno2002PRA}, with the boundary conditions shown in Table \ref{TabBC}.
Here, we used 50 basis splines for each direction in the $\theta$-$\varphi$ hyperangular plane and obtained at least six digits of accuracy for the potentials
$U_{\nu}$ for values of $R$ up to 10$r_0$. As usual, we used a non-uniform grid distribution, concentrating more points near the potential minima
where the channel functions change more drastically. In what follows, we will present the results in units based on the short-range length
scale $r_0$ ---i.e., length is in units of $r_0$ and energy in units of $1/m r_0^2$, where $m$ is the mass of each identical particle.

\subsection{Results}

\begin{figure*}[htbp]
\includegraphics[width=6.9in]{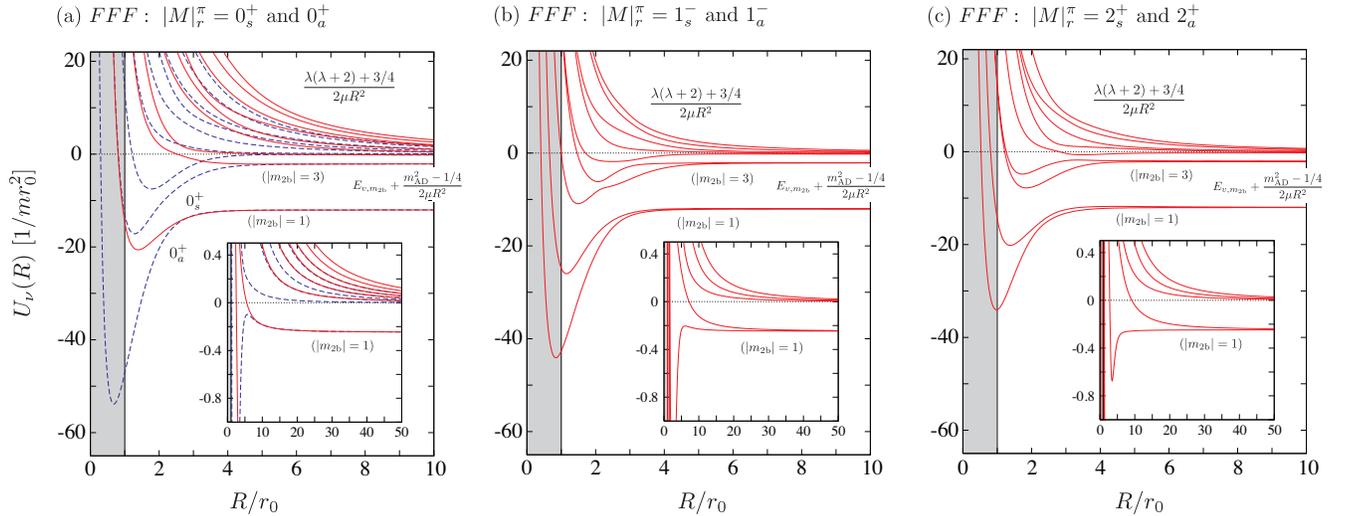}
\caption{(color online) Three-body potentials for three identical fermions in 2D with (a) $|M|^{\pi}_{r}=0^+_s$ and $0^+_a$,
(b) $|M|^{\pi}_{r}=1^-_s$ and $1^-_a$, and (c) $|M|^{\pi}_{r}=2^+_s$ and $2^+_a$ (shaded region indicates the region $R$ in which particles can be found
at distances smaller than $r_0$). Note that, similar to identical bosons, the
$|M|\ne0$ potentials for the symmetric and antisymmetric 2D three-body systems (with respect to reflection) are exactly degenerate,
while for $|M|=0$ they are not.}
\label{FermionsFig}
\end{figure*}

To illustrate the structure of the 2D three-body potentials and gain information about the
collisional properties of the system, we have calculated the three-body potentials $U_{\nu}(R)$ for three identical 
bosons and three identical fermions for various values of $|M|^\pi_r$, with the results shown in Figs. \ref{BosonsFig} and \ref{FermionsFig}. 
For these calculations, we assumed the interatomic interaction to be a pairwise sum of 
two-body interactions as shown in Eq.~(\ref{Int}), where the two-body interaction is given
by the short-range potential 
\begin{eqnarray}
v(r)=D{\rm sech}^2(r/r_{0}). 
\end{eqnarray}
Here, $D$ is the potential depth and $r_{0}$ is the 
range of the interaction. The results in Figs. \ref{BosonsFig} and \ref{FermionsFig} were obtained for 
$D=-30/mr_0^{2}$, such that $v(r)$ supports three two-body bound states with total angular 
momentum $m_{\rm 2b}=0$ ($s$-wave), two with $|m_{\rm 2b}|=1$ ($p$-wave), one with $|m_{\rm 2b}|=2$ ($d$-wave), 
and one with $|m_{\rm 2b}|=3$ ($f$-wave). Note that for identical bosons the only states allowed by symmetry have
$m_{\rm 2b}$ even while identical fermions have $m_{\rm 2b}$ odd.
 
Results for $|M|=0$, $1$ and $2$ are given in Figs.~\ref{BosonsFig} and \ref{FermionsFig} 
labeled by $|M|^{\pi}_{r}$, indicating the parity $\pi=(-)^M$ as well as the reflection 
symmetry $r=s,a$ for symmetric and antisymmetric states, respectively. For states with 
$|M|\neq0$, the symmetric and antisymmetric solutions, $|M|^{\pi}_s$ and $|M|^{\pi}_a$, are exactly degenerate,
which is equivalent to the degeneracy of $+M$ and $-M$. For $|M|=0$ states [see Figs.~\ref{BosonsFig}(a) and \ref{FermionsFig}~(a)], 
however, the $0^+_s$ and $0^+_a$ states are non-degenerate and possess distinct
properties for both bound and scattering states. Such differences, however, will become more clear below once we discuss the asymptotic behavior
($R\gg r_{0}$) of the three-body potentials.
At these distances two configurations for the three-body system are possible: three free particles (three-body continuum channels)
and a diatom plus a free particle (atom-diatom channels). 

\subsection{Three-body continuum channels}

As we saw in Sec. \ref{HypersphericalHarmonics}, the motion of three free particles can be described
in terms of hyperspherical harmonics [see Eqs.~(\ref{EigenValueProb}) and (\ref{Harmonics})]. Therefore, the leading-order behavior 
of the continuum channels is 
\begin{eqnarray} 
W_{\nu}(R) \underset{R\rightarrow\infty}{\longrightarrow}\frac{\lambda(\lambda+2)+3/4}{2\mu R^2},\label{3BCont}
\end{eqnarray}
with the allowed values for $\lambda$ given in Tables \ref{ThreeBosonTab}--\ref{TwoFermionTab} in Sec. \ref{HypersphericalHarmonics}. 
Our numerical results show that $Q_{\nu\nu}$ falls faster than $1/R^2$. 

As mentioned above, $|M|^\pi_s$ and $|M|^{\pi}_a$ states with $M\neq0$ are degenerate.
For $0^+_s$ and $0^+_a$, however, the potentials are not degenerate, and the results in Tables 
\ref{ThreeBosonTab}--\ref{TwoFermionTab} include both symmetry states for $M=0$. For instance, for three identical bosons, the 
$\lambda=0$ solution is symmetric with respect to reflection and is therefore a state of $0^+_s$ 
symmetry. 
The lowest value of $\lambda$ possible for $M=0$ thus occurs for $0^+_s$, and we will see that
$\lambda_{\rm min}$ largely determines the low-energy behavior of scattering observables. For $0^+_a$, $\lambda_{\rm min}=8$.
The opposite scenario holds for identical fermions where $\lambda_{\rm min}=2$ for $0^+_a$ while $\lambda_{\rm min}=6$ for
$0^+_s$.

\subsection{Atom-diatom channels}

Asymptotically, $W_{\nu}$ for atom-diatom channels approaches the energy 
of the two-body bound states, $E_{v,m_{\rm 2b}}$, $v$ being the vibrational quantum number, with a leading $1/R^2$ behavior:
\begin{eqnarray} 
W_{\nu}(R) \underset{R\rightarrow\infty}{\longrightarrow} E_{v,m_{\rm 2b}}+\frac{m_{\rm AD}^2-1/4}{2\mu R^2}.\label{3BAD}
\end{eqnarray} 
Here, $m_{\rm AD}$ is the relative angular momentum between atom and diatom, satisfying $M=m_{\rm 2b}+m_{\rm AD}$. 
For atom-diatom channels, the $-Q_{\nu\nu}/2\mu$ term is proportional to
$1/R^2$, and its inclusion is crucial in order to properly recover $W_{\nu}$ at large distances since it exactly cancels an attractive term in $W_{\nu}$.
[Note that in Figs.~\ref{BosonsFig} and \ref{FermionsFig} we show the numerical results for $U_{\nu}$ but
indicate the asymptotic behavior corresponding to $W_{\nu}$.]

For channels with $m_{\rm 2b}\neq0$, two atom-diatom potentials converge to each threshold
with $m_{\rm AD}=M\mp|m_{\rm 2b}|$ corresponding to the $m_{\rm 2b}=\pm|m_{\rm 2b}|$ states. 
For $m_{\rm 2b}=0$ states, of course, only $m_{\rm AD}=M$ is allowed. These
properties can be seen in the three-body potentials shown in Figs. \ref{BosonsFig}(b) and \ref{BosonsFig}(c) 
and Figs.~\ref{FermionsFig}(b) and \ref{FermionsFig}(c). 
These statements must be qualified for the $0^+_s$ and $0^+_a$ curves in Figs. \ref{BosonsFig}(a) and \ref{FermionsFig}(a).
Specifically, in  Fig. \ref{BosonsFig}(a), two-boson states with $m_{\rm 2b}=0$ are not allowed for $0^+_a$ since no reflection-antisymmetric 
solution can be constructed with $m_{\rm 2b}=0$. Further, for both bosons and fermions with $m_{\rm 2b}\ne0$,
one of the two possible values of $m_{\rm AD}$ is of $0^+_s$ symmetry, and the other is $0^+_a$. 

\section{Threshold laws for inelastic collisions in 2D} \label{ThresholdLaws}

In this section, we determine the threshold laws \cite{esry2001PRA}, i.e., the low-energy dependence, of three-body scattering observables 
in 2D that are relevant for ultracold atoms. We have previously shown \cite{esry2001PRA,dincao2005PRL} that the energy
dependence for inelastic scattering observables can be obtained by a simple WKB analysis and relies mostly on the asymptotic behavior 
of the initial potential relevant for the collision process. In fact, from Eqs.~(\ref{3BCont}) and (\ref{3BAD}) we see that both atom-diatom and continuum 
channels can be written in terms of the familiar 3D centrifugal potential,
\begin{equation}
W_\nu(R)\underset{R\rightarrow\infty}{\longrightarrow} \frac{\ell_{\rm eff}(\ell_{\rm eff}+1)}{2\mu R^2},\label{3DCB}
\end{equation}
with an effective angular momentum $\ell_{\rm eff}$ that depends on either $\lambda$ or $m_{\rm AD}$. 
It is now straightforward to derive the threshold laws.

The energy dependence of scattering observables can be determined \cite{dincao2005PRL} from the observation that inelastic 
transitions occur at distances much smaller than the classical turning point $r_c$.
In this case, the relevant transition probability $|T_{f\leftarrow i}|^2$ is proportional to the probability for 
all three particles to approach to such distances. We can thus approximate $|T_{f\leftarrow i}|^2$ using
the WKB approximation for the tunneling probability from $r_{c}$ to distances comparable to $r_0$,
\begin{align}
&|T_{f\leftarrow i}|^2\propto 
\nonumber\\
&~~~\exp\left[-2\int_{r_{0}}^{r_c}\sqrt{2\mu\left(W_{i}(R)+
\frac{{1}/{4}}{2\mu R^2}-E\right)}dR\right].
\label{TransProb}
\end{align}
In this expression, we have included the Langer correction \cite{berry1972RPP} and will use Eq.~(\ref{3DCB})
for $W_i(R)$. The turning point is then $r_c=(\ell_{\rm eff}+\frac{1}{2})/k$ with $k^2=2\mu E$.
Therefore, in the WKB approximation, the energy 
dependence of the transition probability can be derived from the above integral, leading to
\begin{eqnarray}
|T_{f\leftarrow i}|^2\propto (k r_0)^{2\ell_{\rm eff}+1},\label{Tmatrix}
\end{eqnarray}
with the value of $\ell_{\rm eff}$ determined from the symmetry of the system (through $\lambda$ and $m_{\rm AD}$)
as described in the previous sections. 

We emphasize that our treatment here and the threshold laws in Eq.~(\ref{Tmatrix}) assumes that any corrections to Eq.~(\ref{3DCB}) are
sufficiently short-ranged compared to $R^{-2}$. However, when there exists a weakly bound $m_{\rm 2b}=0$ two-body state, it is not known 
what form the threshold laws will take. Equation (\ref{Tmatrix}) no longer applies since $m_{\rm 2b}=0$ two-body states imply a large
admixture of logarithm-containing terms in the low-energy two-body scattering state \cite{shih2009PRA,kanjilal2006PRA}. 
These, in turn, lead to logarithm-containing terms in $W_{\nu}(R)$ \cite{kartavtsev2006PRA,helfrich2011PRA} that are not sufficiently short-ranged compared to $R^{-2}$. 
For non-resonant interactions, however, our results should be valid. A more detailed study of the resonant case will be the subject of future analysis.

\subsection{Atom-diatom collisions}

In an ultracold mixture of atoms and molecules, the collisional processes primarily responsible for the mixture's
stability are ro-vibrational relaxation,
\begin{eqnarray}
XY(v,m_{\rm 2b})+Z\rightarrow XY(v',m_{\rm 2b}')+Z,\nonumber
\end{eqnarray}
and reactive scattering,
\begin{eqnarray}
XY(v,m_{\rm 2b})+Z\rightarrow XZ(v',m_{\rm 2b}')+Y.\nonumber
\end{eqnarray}
At ultracold temperatures, these reactions only occur if they are exothermic since the collision energy is generally orders of magnitude smaller
than excitation energies. By the same token, the excitation energy released in these collisions as relative kinetic energy of the
fragments is usually sufficient for the fragments to escape the trapping potential, thus leading to loss from the mixture.
The energy dependence of both scattering processes, however, is the same as it only depends on the
properties of the initial collision channel, $XY(v,m_{\rm 2b})+Z$.

For atom-diatom collisions, the initial collision channel is described at large distances by Eq.~(\ref{3BAD}). 
Comparison with Eq.~(\ref{3DCB}) allows us to identify $\ell_{\rm eff}=m_{\rm AD}-1/2$. From Eq.~(\ref{Tmatrix}) and the fact that the atom-diatom
inelastic collision rate is proportional to $|T|^2$, we obtain
\begin{eqnarray}
{K}_{\rm AD}^{(M)} \propto (k_{\rm AD}r_{0})^{2|m_{\rm AD}|},\label{ADRate}
\end{eqnarray}
where $k_{\rm AD}^2=2\mu_{\rm AD}(E-E_{v,m_{\rm 2b}})$ and $\mu_{\rm AD}$ is the atom-diatom reduced mass. 
The above expression is valid for energies smaller than the smallest energy scale in the system \cite{dincao2004PRL}. 
Therefore, if no other two-body state is more weakly bound than the diatom state, 
Eq.~(\ref{ADRate}) should be valid for $k_{\rm AD}^2\ll 2\mu_{\rm AD}|E_{v,m_{\rm 2b}}|$.

Equation (\ref{ADRate}) states that the threshold law is determined by the smallest value of $|m_{\rm AD}|$ allowed.
Although it is not directly determined by identical particle symmetry, symmetry does indirectly influence it
via $m_{\rm 2b}$ in $m_{\rm AD}=M\pm|m_{\rm 2b}|$.
As we saw previously, for a given $M$ and $m_{\rm 2b}=0$ ($s$-wave) only one value
for $m_{\rm AD}$ is allowed, $m_{\rm AD}=M$. For $m_{\rm 2b}\ne0$ two values for $m_{\rm AD}$ are allowed. 
We are interested, however, only on the lowest value for $|m_{\rm AD}|$, $|m_{\rm AD}|=||M|-|m_{\rm 2b}||$,
since it gives the dominant contribution to $K_{\rm AD}$ as $k_{\rm AD}\rightarrow0$. 

In Table \ref{ThresholdLawsTab} we summarize the threshold laws for the atom-diatom inelastic rate for 
all combinations of identical bosons and fermions for the lowest few 
values of $|M|$. 
Note that we use the 3D notation for angular momentum, i.e., $m_{\rm 2b}=0,1,2,...$ are represented by $s,p,d,...$.
Note also that for $BBX$ systems, even values of $m_{\rm 2b}$ can be attributed either to a $BB$ or $BX$ molecule, while if $m_{\rm 2b}$ is odd
it only can be attributed to a $BX$ diatom. On the other hand, for $FFX$ systems, even $m_{\rm 2b}$ can only be
attributed to an $FX$ diatom, while odd $m_{\rm 2b}$ can be either an $FF$ or $FX$ diatom. Based on these results,
we can conclude that for any system the dominant partial wave ($M$) contribution to $K_{\rm AD}$
(see underlined results in Table~\ref{ThresholdLawsTab}) will be $|M|=|m_{\rm 2b}|$ and that $K_{\rm AD}$ 
is constant as $k_{\rm AD}\rightarrow0$, even for identical fermions.

\begin{table}[htbp]
\begin{ruledtabular}
\caption{Summary of the threshold laws for inelastic three-body processes in 2D for the lowest few values of $|M|^{\pi}_{r}$.
Results are for atom-diatom inelastic rates ($K_{\rm AD}$), three-body recombination ($K_{3}$) and collision-induced dissociation ($D_{3}$) 
for systems of three identical bosons ($BBB$), three identical fermions ($FFF$), two identical bosons ($BBX$) and two identical 
fermions ($FFX$), for both symmetric- and antisymmetric-reflection symmetries. The dominant partial wave contribution ($M$) is underlined for each system. }
\begin{tabular}{ccccccc}
         & $|M|^\pi_r$& $|m_{\rm 2b}|(|m_{\rm AD}|)$ & $K^{(M)}_{\rm AD}$ & $\lambda_{\rm min}$ & $K_3^{(M)}$ & $D_3^{(M)}$ \\ \hline 
$BBB$   
         & $0^+_s$ &     $s$(0),$d$(2)  & $\underline{k_{\rm AD}^0},k_{\rm AD}^4$     & 0  & $\underline{k^0}$  & $\underline{k^2}$ \\
         & $0^+_a$ &     $d$(2),$g$(4)  & $k_{\rm AD}^4,k_{\rm AD}^8$     & 8  & $k^{16}$  & $k^{18}$ \\
         & $1^-_{s/a}$ & $s$(1),$d$(1)  & $k_{\rm AD}^2,k_{\rm AD}^2$     & 3  &  $k^6$  & $k^8$ \\
         & $2^+_{s/a}$ & $s$(2),$d$(0)  & $k_{\rm AD}^{4},\underline{k_{\rm AD}^{0}}$  & 2  & $k^4$  &  $k^6$ \\ \hline
$BBX$   
         & $0^+_s$ &     $s$(0),$p$(1),$d$(2)  & $\underline{k_{\rm AD}^0},k_{\rm AD}^{2},k_{\rm AD}^4$      & 0  & $\underline{k^0}$   & $\underline{k^2}$ \\
         & $0^+_a$ &     $p$(1),$d$(2),$f$(3)  & $k_{\rm AD}^{2},k_{\rm AD}^4,k_{\rm AD}^8$      &  4 &$k^{8}$   & $k^{10}$  \\
         & $1^-_{s/a}$ & $s$(1),$p$(0),$d$(1)  & $k_{\rm AD}^2,\underline{k_{\rm AD}^0},k_{\rm AD}^{2}$      & 1  &  $k^2$  & $k^4$  \\
         & $2^+_{s/a}$ & $s$(2),$p$(1),$d$(0)  & $k_{\rm AD}^{4},k_{\rm AD}^{2},\underline{k_{\rm AD}^{0}}$   & 2  &  $k^4$  &  $k^6$ \\ \hline
$FFF$
         & $0^+_s$ &     $p$(1),$f$(3)  & $k_{\rm AD}^2,k_{\rm AD}^6$  & 6  & $k^{12}$   & $k^{14}$  \\
         & $0^+_a$ &     $p$(1),$f$(3)  & $k_{\rm AD}^2,k_{\rm AD}^6$  & 2  & $\underline{k^4}$   & $\underline{k^6}$  \\
         & $1^-_{s/a}$ & $p$(0),$f$(2)  & $\underline{k_{\rm AD}^0},k_{\rm AD}^4$  &  3 & $k^6$   & $k^8$  \\
         & $2^+_{s/a}$ & $p$(1),$f$(1)  & $k_{\rm AD}^2,k_{\rm AD}^2$  &  4 & $k^8$   & $k^{10}$  \\ \hline
$FFX$   
         & $0^+_s$ &     $s$(0),$p$(1),$d$(2)  & $\underline{k_{\rm AD}^0},k_{\rm AD}^{2},k_{\rm AD}^4$     & 2  & $k^4$  & $k^6$  \\
         & $0^+_a$ &     $p$(1),$d$(2),$f$(3)  & $k_{\rm AD}^{2},k_{\rm AD}^4,k_{\rm AD}^8$     & 2  & $k^{4}$   & $k^{6}$  \\
         & $1^-_{s/a}$ & $s$(1),$p$(0),$d$(1)  & $k_{\rm AD}^2,\underline{k_{\rm AD}^0},k_{\rm AD}^{2}$     &  1 & $\underline{k^2}$   & $\underline{k^4}$  \\
         & $2^+_{s/a}$ & $s$(2),$p$(1),$d$(0)  & $k_{\rm AD}^{4},k_{\rm AD}^{2},\underline{k_{\rm AD}^{0}}$  &  2 & $k^4$   & $k^6$  
\end{tabular}
\label{ThresholdLawsTab}
\end{ruledtabular}
\end{table}

\subsection{Three-body recombination}

In ultracold atomic gases, three-body recombination, 
\begin{eqnarray}
X+Y+Z\rightarrow XY+Z, 
\end{eqnarray}
is often the major atom-loss mechanism since the atomic states are typically chosen to eliminate two-body collisional losses.
The atom and diatom produced have large kinetic energy and are thus lost from the trap. In general, calculating
three-body recombination rates requires the infinity of initial continuum channels [Eq.~(\ref{3BCont})], making the 
calculations extremely challenging. 
Fortunately, at ultracold collision energies, the lowest continuum channel provides the dominant contribution. 
This simplification allows us to apply the WKB approach above by 
identifying $\ell_{\rm eff}=\lambda+1/2$, leading to 
\begin{eqnarray}
{K}_{3}^{(M)} \propto k^{2\lambda}r_{0}^{2\lambda+2},\label{K3rate}
\end{eqnarray}
for the three-body recombination rate, $K_{3}\propto|T|^2/k^2$, where $k^{2}=2\mu E$.
This result is also expected to be valid for energies much smaller than any other energy scale in the system 
\cite{dincao2004PRL}. Therefore, Eq.~(\ref{K3rate}) shows that for a given $M$ the dominant channel is determined 
by $\lambda=\lambda_{\rm min}$. 

Table \ref{ThresholdLawsTab} includes the three-body recombination threshold laws for all combinations of identical particles.
The dominant partial wave contribution--- i.e., the one that has the lowest value 
for $\lambda_{\rm min}$--- is underlined in Table \ref{ThresholdLawsTab}. 
For instance, for bosonic systems $BBB$ and $BBX$, $M=0$ is dominant with $\lambda_{\rm min}=0$, implying that $K_{3}$ is constant for $E\rightarrow0$. 
For $FFF$ systems, the dominant contribution is still $M=0$, but with $\lambda_{\rm min}=2$ giving $K_3\propto k^4$. 
For $FFX$ systems, $M=1$ dominantes with $\lambda_{\rm min}=1$ for $K_{3}\propto k^2$. 

We note that for three identical bosons
near an $m_{\rm 2b}=0$ resonance, Ref.~\cite{helfrich2011PRA} found $K_{3}$ to vanish in the limit of $E\rightarrow0$. 
The most likely explanation of this disagreement with the present analysis is that the logarithmic terms in $W_{\nu}(R)$
mentioned above do indeed have a dramatic effect on the threshold laws.

\subsection{Collision-induced dissociation}

The time-reverse of three-body recombination, collision-induced dissociation,
\begin{eqnarray}
XY+Z\rightarrow X+Y+Z, 
\end{eqnarray}
is only allowed if the collision energy is greater than the diatom binding energy such that the dissociation channels 
(three-body continuum channels) are energetically accessible. Therefore, at ultracold temperatures only dissociation of weakly bound molecules is possible.
Because the collision-induced dissociation rate $D_3$ behaves as $D_{3}\propto(kr_0)^{2\lambda+2}(k_{\rm AD}r_0)^{2|m_{\rm AD}|}$, it is the final three-body
continuum channel that determines the threshold law since $k_{\rm AD}$ is finite at the breakup threshold where $k=0$. The threshold law thus simplifies to
\begin{eqnarray}
{D}_{3}^{(M)}\propto (k r_{0})^{2\lambda+2}.
\end{eqnarray}
We note that the energy dependence for $D_{3}$ differs from the one for $K_{3}$ by a factor $k^2$ due to the difference in phase-space factors from the
reversed roles of the initial and final states. The $D_{3}$ threshold laws are also summarized in Table \ref{ThresholdLawsTab}.

\section{Summary} \label{Summary}

We have explored three-body systems in two dimensions using the adiabatic hyperspherical representation.
We derived symmetry properties and boundary conditions for all permutation symmetries, establishing
an efficient numerical approach for solving three-body problems in 2D. From explicit numerical examples, we
demonstrated the asymptotic behavior of the three-body potentials and illustrated the topology of such potentials.
From this analysis, complemented by our symmetry considerations, we were able to determine the threshold laws
for atom-diatom inelastic collisions, as well as three-body recombination and collision-induced dissociation
for various partial waves and symmetries. These results can be used for determining the expected collisional
behavior and stability of ultracold atomic and molecular gases in two dimensions. The hyperspherical formalism 
we outline in this work is capable of treating three-body systems in which the two-body interactions
can support deeply bound states as well as weakly bound states. Thus, in contrast to the formalisms in which a zero-range model potential is used,
our approach is suitable for studying more realistic systems where the finite range aspect of the interatomic
interactions plays an important rule.

\acknowledgments

We thank to Fatima Anis for helpful comments in the earlier stages of this project. This work was
was supported by the U.S. National Science Foundation and the AFOSR-MURI.


\appendix

\section{Effects of symmetry operators} \label{Permutations}

\begin{figure}[htbp]
\includegraphics[scale=0.4]{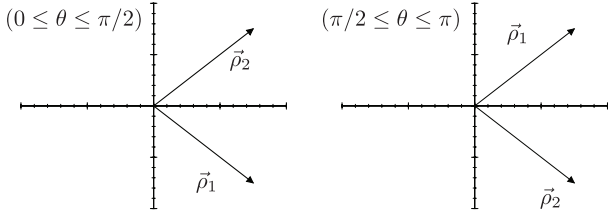} 
\caption{Relation of the coordinate $\theta$ to the relative positions of the
Jacobi vectors in the body frame.}
\label{ThetaFig}
\end{figure}

To understand the effects of the symmetry operators, it is simplest to draw pictures
of the mass-weighted Jacobi vectors in the center of mass frame.  It is also necessary
to understand the role of the coordinate $\theta$.  The essential point is illustrated
in Fig.~\ref{ThetaFig} which shows that the value of $\theta$ indicates the relative
positions of the two Jacobi vectors.

The general scheme will then be to draw the Jacobi vectors, then draw the vectors resulting
from the symmetry operation.  The changes to the coordinates will be inferred from comparing the figures.

Because we do not change the moments of inertia with these symmetry operations, 
the body frame $x$ and $y$ axes can at most be inverted since they are defined 
from the Jacobi vectors.  Consequently, $\gamma$ can only be changed by 0 or $\pi$.
It helps to know the moments of inertia:
\begin{align}
I_{xx} & = R^2 \sin^2\left(\frac{\theta}{2}-\frac{\pi}{4}\right) \\
I_{yy} & = R^2 \cos^2\left(\frac{\theta}{2}-\frac{\pi}{4}\right).
\end{align}
For the same reason, $\theta$ can only be changed to $\pi-\theta$, if it is
changed at all.  This fact has the convenient consequence that we can pick
a particular $\theta$ to work with and know that our results work for all
$\theta$.  Thus, we will pick $\theta$=0 (equilateral triangle) so that
$\vec{\rho}_1$ is orthogonal to $\vec{\rho}_2$. 
Given the coordinates' dependence on $\theta$ and $\varphi$ [see Eq.~(\ref{Coords})], in what follows it is 
simpler to work with $\theta'=\left(\frac{\theta}{2}-\frac{\pi}{4}\right)$ and $\varphi'=\left(\frac{\varphi}{2}+\frac{\pi}{6}\right)$.  
The coordinate $\varphi'$ just measures the angle of $\vec{\rho}_2$ from the $x$ axis and takes on values between $\frac{\pi}{6}$ 
and $\frac{7\pi}{6}$. Nevertheless, our results will finally be expressed in terms of $\theta$ and $\varphi$.
Note that we found we had to be careful to split each operation up over two intervals in $\varphi$ \cite{johnson1983JCP}.  
Not too surprisingly, it turns out that even though the coordinates are affected differently in the two intervals, 
the functions that depend on them are not. This property allows us to obtain the single expression for each operator shown in Eqs.~(\ref{Pfirst})--(\ref{Plast}),
valid over the whole range of $\varphi$.

\subsection{Parity}

\begin{figure}[htbp]
\centerline{\includegraphics[scale=0.50]{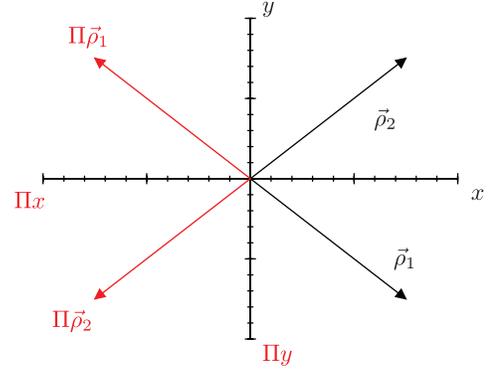}}
\caption{ Jacobi vectors before (black) and after (red) the parity operation, $\Pi$.}
\label{ParityFig}
\end{figure}

The parity operation has the following effects on the mass-scaled Jacobi vectors 
\begin{align}
\Pi (\vec{\rho}_1,\vec{\rho}_2) =  (-\vec{\rho}_1,-\vec{\rho}_2). 
\end{align}
This operation is illustrated in Fig.~\ref{ParityFig}. Now, using Eq.~(\ref{Coords}) of the main text, we find that the hyperspherical coordinates 
are affected by parity operation as
\begin{equation}
\Pi (\theta,\varphi,\gamma) = (\theta,\varphi,\pi+\gamma).
\end{equation}
As a consequence, it is easy to determine that the hyperspherical harmonics [Eq.~(\ref{Harmonics})] are affected by parity as
\begin{equation}
\Pi
Y_{\omega M}^\lambda(\Omega) = (-)^MY_{\omega M}^\lambda(\Omega),
\end{equation}
where $\Omega\equiv\{\theta,\varphi,\gamma\}.$

\subsection{Permutation: $P_{12}$}

Choosing $\vec{\rho}_1$ as the Jacobi vector connecting particles 1 and 2, $P_{12}$ has the following effect
\begin{align}
P_{12} (\vec{\rho}_1,\vec{\rho}_2) = (-\vec{\rho}_1,\vec{\rho}_2).
\end{align}
\begin{figure}[htbp]
\centerline{\includegraphics[scale=0.50]{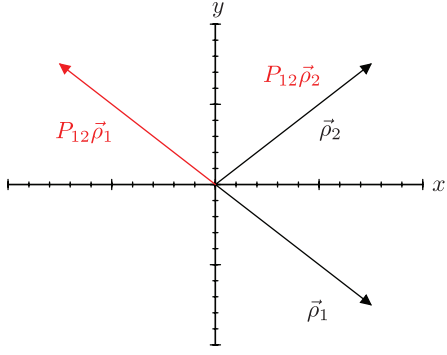}}
\caption{Jacobi vectors before (black) and after (red) the $P_{12}$ permutation operation.}
\label{P12Fig}
\end{figure}

First, since the relative positions of $\vec{\rho}_1$ and $\vec{\rho}_2$ have changed,
we know $P_{12}\theta'=-\theta'$ ($P_{12}\theta=\pi-\theta$). Second,
since $\varphi'$ has the same range as $\varphi$, we must make sure that both stay
within this range.  After some head scratching and careful drawing, we find the following:
\begin{equation}
P_{12}(\theta',\varphi',\gamma) = 
\begin{cases}
(-\theta',\pi-\varphi',\pi+\gamma), &{\rm for}~ \mbox{$\frac{\pi}{6}$$\le$$\varphi'$$\le$$\frac{5\pi}{6}$} \\
(-\theta',2\pi-\varphi',\gamma), &{\rm for}~ \mbox{$\frac{5\pi}{6}$$\le$$\varphi'$$\le$$\frac{7\pi}{6}$},
\end{cases}
\end{equation}
or
\begin{equation}
P_{12}(\theta,\varphi,\gamma) = 
\begin{cases}
(\pi-\theta,\frac{4\pi}{3}-\varphi,\pi+\gamma), &{\rm for}~ \mbox{$0$$\le$$\varphi$$\le$$\frac{4\pi}{3}$} \\
(\pi-\theta,\frac{10\pi}{3}-\varphi,\gamma), &{\rm for}~ \mbox{$\frac{4\pi}{3}$$\le$$\varphi$$\le$$ 2\pi$}.
\end{cases}
\end{equation}

Now, from Eq.~(\ref{Harmonics}) and using $d_{mm'}^\ell (\pi-\theta) = (-)^{\ell+m'}d_{-mm'}^\ell(\theta)$ and Eq.~(\ref{OneToOne}), 
one can show that the effect of the permutation $P_{12}$ on the hyperspherical harmonics does not depend on the range in $\varphi$
and is given by
\begin{eqnarray}
P_{12} Y_{\omega M}^\lambda (\Omega) = (-)^\frac{3M+\lambda}{2} e^{i\omega\frac{2\pi}{3}}
       Y_{-\omega M}^\lambda (\Omega).
\end{eqnarray}

\subsection{Permutation: $P_{23}$}

\begin{figure}[htbp]
\centerline{\includegraphics[scale=0.50]{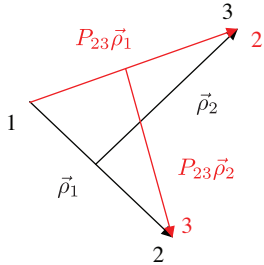}}
\caption{Jacobi vectors before (black) and after (red) the $P_{23}$ permutation operation.}
\label{P23Fig}
\end{figure}

For this operation, we could write out the explicit changes in $\vec{\rho}_1$ and $\vec{\rho}_2$,
but we believe it is easier to just draw the vectors illustrating the effect of $P_{23}$ as shown in Fig.~\ref{P23Fig}.

We should now make a drawing like Fig.~\ref{P12Fig}, but hopefully the idea is getting clear.
The key is to realize that since the particles are in an equilateral configuration, the
angle between $\Pi\vec{\rho}_1$ and $\vec{\rho}_1$ is $\frac{\pi}{3}$.  We similarly know
all of the other relative angles --- and they are simple --- which is why we chose to work with $\theta=0$. We find
\begin{equation}
P_{23}(\theta',\varphi',\gamma) = 
\begin{cases}
(-\theta',\frac{2\pi}{3}-\varphi',\gamma), &{\rm for}~ \mbox{$\frac{\pi}{6}$$\le$$\varphi'$$\le$$\frac{\pi}{2}$} \\
(-\theta',\frac{5\pi}{3}-\varphi',\pi+\gamma), &{\rm for}~ \mbox{$\frac{\pi}{2}$$\le$$\varphi'$$\le$$\frac{7\pi}{6}$},
\end{cases}
\end{equation}
or
\begin{equation}
P_{23}(\theta,\varphi,\gamma) = 
\begin{cases}
(\pi-\theta,\frac{2\pi}{3}-\varphi,\gamma), &{\rm for}~ \mbox{$0$$\le$$\varphi$$\le$$\frac{2\pi}{3}$} \\
(\pi-\theta,\frac{8\pi}{3}-\varphi\pi+\gamma), &{\rm for}~ \mbox{$\frac{2\pi}{3}$$\le$$\varphi$$\le$$ 2\pi$}.\label{P23a}
\end{cases}
\end{equation}

Similar to $P_{12}$, one can show that the effect of $P_{23}$ does not depend on the range in $\varphi$ and is given by
\begin{eqnarray}
P_{23} Y_{\omega M}^\lambda (\Omega) = (-)^\frac{M+\lambda}{2} e^{i\omega\frac{\pi}{3}}
       Y_{-\omega M}^\lambda (\Omega).\label{P23b}
\end{eqnarray}

\subsection{Permutations $P_{31}$, $P_{12}P_{23}$ and $P_{12}P_{31}$}

Making drawings like Fig.~\ref{P23Fig} and Fig.~\ref{P12Fig}, it is possible to derive the effect of the permutations
$P_{31}$, $P_{12}P_{23}$ and $P_{12}P_{31}$. Note that these effects could be worked out from the 
$P_{12}$ and $P_{23}$, but because of the different ranges for $\varphi$, this approach would be complicated.
We will forego more figures, however, and simply state the results for the remaining permutations and corresponding actions on the
hyperspherical harmonics.

For $P_{31}$ the result is particularly simple (the range of $\varphi$ is not split),
\begin{equation} 
P_{31}(\theta,\varphi,\gamma) = (\pi-\theta,2\pi-\varphi,\gamma),\label{P31}
\end{equation}
which leads to
\begin{equation}
P_{31} Y_{\omega M}^\lambda (\Omega) = (-)^\frac{3M+\lambda}{2}
       Y_{-\omega M}^\lambda (\Omega).
\end{equation}
For $P_{12}P_{23}$ and $P_{12}P_{31}$, we obtain 
\begin{equation}
P_{12}P_{23}(\theta,\varphi,\gamma) = 
\begin{cases}
(\theta,\frac{2\pi}{3}+\varphi,\pi+\gamma), &{\rm for}~ \mbox{$0$$\le$$\varphi$$\le$$\frac{4\pi}{3}$} \\
(\theta,-\frac{4\pi}{3}+\varphi,\gamma), &{\rm for}~ \mbox{$\frac{4\pi}{3}$$\le$$\varphi$$\le$$ 2\pi$}.
\end{cases}
\end{equation}
and
\begin{equation}
P_{12}P_{31}(\theta,\varphi,\gamma) = 
\begin{cases}
(\theta,\frac{4\pi}{3}+\varphi,\gamma), &{\rm for}~ \mbox{$0$$\le$$\varphi$$\le$$\frac{2\pi}{3}$} \\
(\theta,-\frac{2\pi}{3}+\varphi,\pi+\gamma), &{\rm for}~ \mbox{$\frac{2\pi}{3}$$\le$$\varphi$$\le$$2\pi$}.
\end{cases}
\end{equation}
These, therefore, lead to
\begin{align}
P_{12}P_{23} Y_{\omega M}^\lambda (\Omega) &= (-)^M e^{i\omega\frac{\pi}{3}}
       Y_{\omega M}^\lambda (\Omega), \\
P_{12}P_{31} Y_{\omega M}^\lambda (\Omega) &= e^{i\omega\frac{2\pi}{3}}
       Y_{\omega M}^\lambda (\Omega).
\end{align}

\end{document}